\documentclass{aa}
\usepackage{natbib}
\usepackage{amssymb}
\usepackage{amsmath}
\usepackage{graphicx}
\usepackage[dvips]{rotating}

\def\mm{\hbox{mm}}

\def\submm{\mbox{(sub-)mm}~}
\begin{document}
\title{Correlation between grain growth and disk geometry in \mbox{Herbig Ae/Be systems}}
\titlerunning{Grain growth in HAEBE disks.}
\authorrunning{Acke et al.}
\author{B.~Acke\inst{1} \and M.E.~van~den~Ancker\inst{2} \and C.P.~Dullemond\inst{3} \and R.~van~Boekel\inst{4} \and L.B.F.M.~Waters\inst{1,4}}
\institute{Instituut voor Sterrenkunde, Katholieke Universiteit Leuven, Celestijnenlaan 200B, 
3001 Leuven, Belgium\\
\email{Bram.Acke@ster.kuleuven.ac.be}
\and
European Southern Observatory, Karl-Schwarzschild Strasse 2, D-85748
Garching bei M\"unchen, Germany
\and
Max-Planck-Institut f\"ur Astrophysik, Karl-Schwarzschildstrasse 1,
Postfach 1317, D-85748 Garching bei M\"unchen, Germany
\and
Astronomical Institute ``Anton Pannekoek'', University of Amsterdam, Kruislaan 403, NL-1098 SJ Amsterdam, The Netherlands
}
\date{DRAFT, \today}


\abstract{We have calculated the \submm spectral indices of 26
Herbig Ae/Be stars, for which we can determine the infrared spectral
energy distribution (SED). We find a clear correlation between the
strength of the ratio of the near- to mid-infrared excess of these sources, and the
slope of the \submm energy distribution. Based on earlier multi-dimensional
modeling of disks around Herbig Ae stars, we interpret this as a
correlation between the geometry of the disk (flared or self-shadowed)
and the size of the grains: self-shadowed disks have, on average, larger
grains than their flared counterparts. These data suggest that the geometry of a
young stellar disk evolves from flared to self-shadowed.
   \keywords{Circumstellar matter -- Planetary systems: protoplanetary
             disks -- Stars: pre-main-sequence} 
}

\maketitle


\section{Introduction}

Herbig Ae/Be (HAEBE) stars are the somewhat
more massive analogues of the T Tauri stars, which are low-mass
young-stellar objects. The spectral energy distribution (SED) of HAEBE stars is characterized
by the presence of an infrared (IR) flux excess, due to circumstellar dust and
gas. The geometry of this circumstellar matter is believed to be
disk-like (e.g. Mannings \& Sargent 1997, 2000; Fuente et al. 2003).

Meeus et al. 2001 (henceforth M01) classified 14 isolated
HAEBE sample stars into two groups, based on the shape of the SED. 
\textit{Group I} contains the sources in which a rising mid-IR
(20--100~$\mu$m) flux
excess is observed; these sources have an SED that can be fitted with a
power-law and a black-body continuum. \textit{Group II} sources have
more modest mid-IR excesses; their IR SEDs can be reconstructed by a power-law
only. M01 suggested phenomologically that the difference between the
two groups is related to the disk geometry.

Dullemond 2002 (henceforth D02) and Dullemond \& Dominik 2004
(henceforth DD04) have modelled young stellar disks with a self-consistent 
model based on 2-D radiative transfer coupled to the equation of
vertical hydrostatics. The model consists of a disk with an inner hole
($\sim$0.5 AU), a puffed-up inner rim and an outer part. The outer
part of the disk can be flared (as in Chiang \& Goldreich 1997), but
can also lie entirely in the shade of the inner rim. The SEDs of
flared disks  
display a strong mid-IR flux excess, while self-shadowed disks have a
much steeper slope towards long wavelengths. D02 can explain the
difference in SED shape in HAEBEs (as expressed by the classification
of M01) as the result of a different disk
geometry; flared-disks systems have group I SEDs, sources with
self-shadowed disks can be linked to group II objects. 

In this paper we calculate the spectral index of the
\submm (350--2700~$\mu$m) SED for a sample of 26 HAEBE stars. We look for a
connection between this index and the geometry of the circumstellar
disk. In Section~\ref{OandA} we 
classify the sample sources and present the computed \submm spectral
indices. Interpretations of the results obtained in the latter section are
given in Section~\ref{interp}. We summarize our conclusions in the final
Section~\ref{concl}.

\section{Observations and Analysis \label{OandA}}

\begin{figure*}
\rotatebox{0}{\resizebox{7in}{!}{\includegraphics{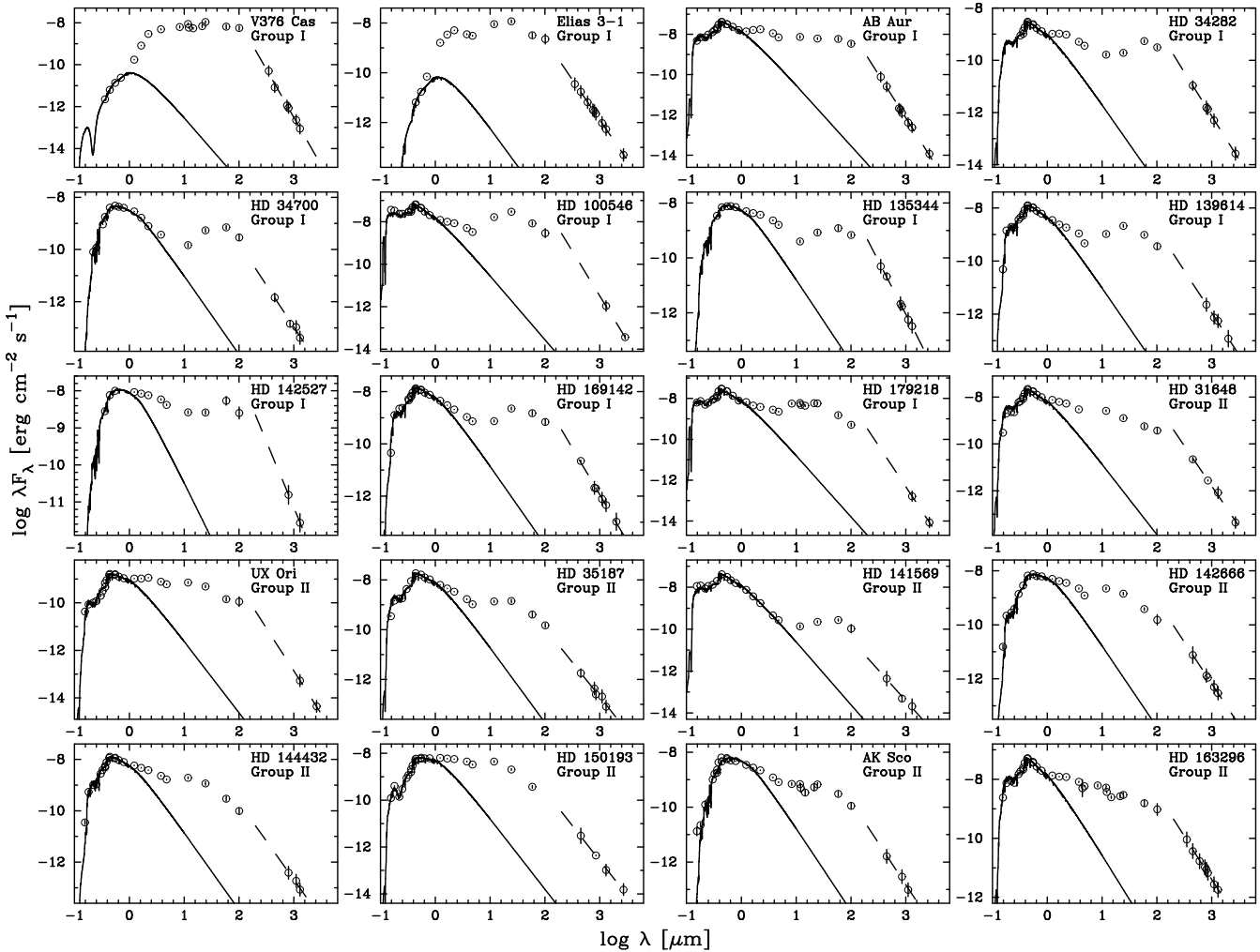}}}
\caption{ The SEDs of the sample sources with sufficient data 
to determine $n$. The circles indicate photometric measurements with
error bars. The full line represents a reddened Kurucz model
corresponding to the photosphere of the central star. The model was
fitted to the optical photometric data. The \submm slope
of the SED is given by the dashed line.}
\label{seds.ps}
\end{figure*}

\begin{figure}
\rotatebox{0}{\resizebox{3.5in}{!}{\includegraphics{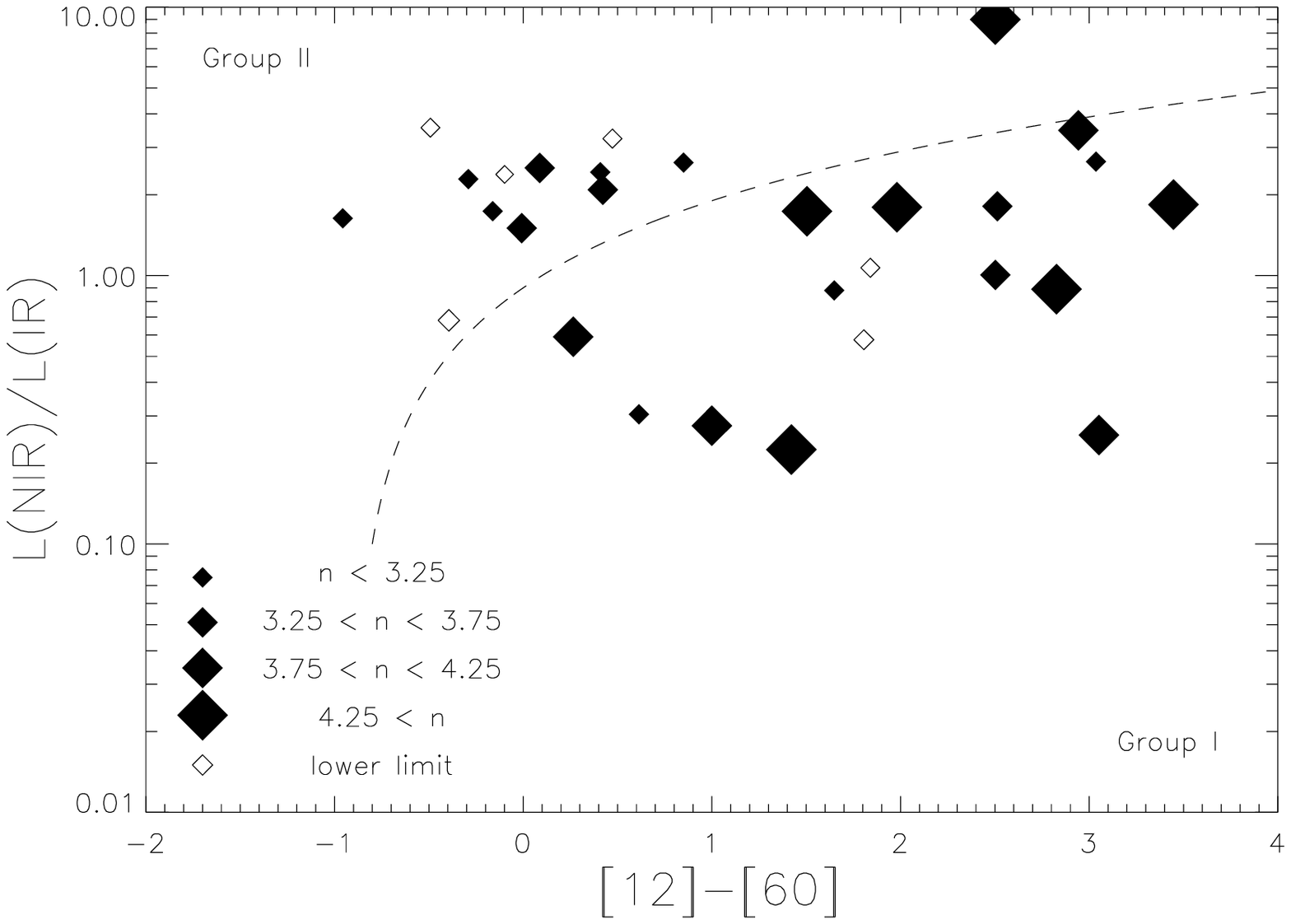}}}
\caption{ Diagram based on van Boekel et al. 2003. The ratio 
$L_{\mathrm{NIR}}/L_{\mathrm{IR}}$ is plotted versus the non
color-corrected $IRAS$ \mbox{[12]-[60]} color. The dashed line empirically
separates the group II (left) and the group I (right) sources. The
size of the plot symbol is proportional to the \submm spectral index
$n$. Open plot symbols indicate lower limits.}
\label{plotroy.ps}
\end{figure}

The sources were selected based on Table 1 and 2 in the latest catalog
of HAEBE stars (Th\'e et al. 1994). We enlarged our list of
HAEBEs with objects from the sample studied by Malfait et al. 1998
that satisfy the criteria postulated by Th\'e et al. We excluded sources that
have the amorphous 10 micron silicate feature in absorption (Acke \&
van den Ancker 2004; henceforth AV04), since these sources are
probably in an earlier evolutionary stage and thus highly
embedded. Early-B stars are excluded as well, because the disk-like
nature of the circumstellar matter in these sources is still a subject of debate
(e.g. see review by Natta et al. 2000 on HAEBE disks in the book
\textit{Protostars and Planets IV}). For the remaining set of stars,
we gathered broad-band $J$, $H$, $K$, $L$ and $M$ photometry and
$IRAS$ 12, 25 and 60 micron measurements in the literature. For
26 of the stars for which this IR photometry was available, also
\submm observations were present in the literature. Our final sample
consists of these sources. 

At \submm wavelengths, one observes the Rayleigh-Jeans part
of the SED of the cold grains. Grains that are large compared to the wavelength at
which they radiate, have a spectrum $\lambda \mathcal{F_{\lambda}}
\propto \lambda^{-n}$ with spectral index $n=3$ (black body). When
the grains are smaller, $n$ is larger, if the dust is
optically thin at these wavelengths. Hence the \submm spectral index
can be interpreted in terms of the grains size distribution of the
cold dust in the circumstellar disk (Sylvester et al. 1996,
van den Ancker et al. 2000, M01). Nevertheless, this should be done
with some caution, as we will discuss in the next section.
We compute the \submm spectral index fitting a power law to the
photometry between 350 and 2700~$\mu$m. In case only one \submm photometric point was
available, we determined a lower limit to the \submm spectral index
using the \submm data point and the $IRAS$ 100 micron measurement,
For BF Ori, we used the $IRAS$ 60 micron measurement, since the $IRAS$ 100
micron measurement resulted in an upper limit in this case.

We note that the fluxes used to compute $n$ come from a variety of literature
sources, both single-dish and interferometric. However, the vast majority of
measurements come from single-dish instruments. There are only four sample
sources (V376 Cas, Elias 3-1, HD 34282 and UX Ori) in which
interferometric data are used. The latter two sources (HD 34282 and UX
Ori) appear rather isolated (Malfait et al 1998). Therefore
the interferometric fluxes are expected to be identical to those
obtained with single-dish measurements. For the two sources with
possible envelope emission (V376 Cas and Elias 3-1; Henning et
al. 1998, di Francesco et al. 1997) for which we have mixed
interferometric and single-dish measurements, we do not note any
systematic differences between the two. In Fig.~\ref{seds.ps} the SEDs
of the sample sources with sufficient \submm photometry to determine
$n$ are presented.

The IR SED of the sample sources can be characterized
with two quantities: the ratio of the near-IR luminosity $L_{\mathrm{NIR}}$
(derived from the broad-band $J$, $H$, $K$, $L$ and $M$ photometry) and the
mid-IR luminosity $L_{\mathrm{IR}}$ (the corresponding quantity based on $IRAS$
12, 25 and 60 micron measurements), and the non-color corrected $IRAS$
[12]-[60] color. Both parameters compare the near-IR photometric data 
to the mid-IR $IRAS$ measurements (van Boekel et al. 2003;
Dullemond et al. 2003). The near-IR SED is expected to 
be similar for all HAEBEs (Natta et al. 2001), while the major
differences occur in the mid-IR SED. The luminosity ratio
$L_{\mathrm{NIR}}/L_{\mathrm{IR}}$ represents the strength of the
near-IR compared to the mid-IR excess, which is lower for group I than
group II sources. The shape of the mid-IR SED of a group I source is
``double-peaked'' compared to the SED of a group II member. The $IRAS$
[12]-[60] color index provides a quantitative measure for this
difference in SED shape. Group I sources are redder than their group
II counterparts. 

Following van Boekel et al. 2003, we use these quantities to
distinguish between group I and II in the classification of M01. We
plotted all sample stars in Fig.~\ref{plotroy.ps}. The dashed line in
the diagram represents $L_{\mathrm{NIR}}/L_{\mathrm{IR}} = ([12]-[60])
+ 0.9$, which empirically provides the best separation between the two
groups. The sources defined by M01 as group I and group II do indeed
occupy the correct location in this diagram using this separating line.
The diagram provides a quantitative alternative to the
subjective classification method of M01.
In Fig.~\ref{plotroy.ps}, the sizes of the plotting symbols are
proportional to the \submm spectral index.

The computed \submm spectral indices $n$ of all sample stars are listed
in Table~\ref{grIorIIvsn}. The error on the spectral index
  $\sigma_n$ reflects the dispersion on the fitted line.
Note that the values of $n$ show
significant variations from source to source. The \submm spectral
indices range from the default ISM value of
optically thin dust $n \approx 5$ (Miyake \& Nakagawa 1993) to the
optically thin black-body value $n=3$. The average values for
group I and group II are also indicated in
Table~\ref{grIorIIvsn}. The error on the average is
given by $(\sum_j 1/\sigma_{n,j})^{-1}$ in which the $\sigma_{n,j}$
represent the errors on the individual $n$-values in the group.

\begin{sidewaystable*}
\caption{ The \submm photometry and spectral index of our sample
sources. The sources are sorted by group. The photometry gathered
from the literature is listed in columns (3) to (10). The unit is
Jy. Between brackets, the observing wavelength in micron is given. In
column (11), the disk mass estimated based on the 1300 micron flux is
presented. The \submm spectral index $n$ and error $\sigma_{n}$ are
given in column (12) and (13) respectively. The references are
indicated in column (14). The mean spectral index of each group is
presented as well. $^\bigstar$ These stars are UX Orionis objects
according to the definition of Dullemond et al. 2003. Note that the
mean spectral index of these objects (3.13$\pm$0.03) is in good
agreement with their believed group II nature (Dullemond et
al. 2003). $^\flat$ Measurement at 600~$\mu$m: 1.3$\pm$0.1 Jy for Elias 3-1
and 3.38$\pm$0.09 Jy for HD 163296. $^\sharp$: average of measurements by
Ref.~15 and 17. $^\dagger$ Based on the 1100 micron flux. $^a$
Measurement at 2~\mm; $^b$ measurement at 2.6~\mm; $^c$ measurement
at 2.9~\mm. 
\label{grIorIIvsn}}	  
\begin{center}
\tabcolsep0.15cm
\scriptsize
\begin{tabular}{|c|c|cccccccc|c|c|c|c|}
 \hline
 \multicolumn{14}{|c|}{\bf \submm photometry and
 spectral index}\\ \hline \hline
 \multicolumn{1}{|c|}{Object} &
 \multicolumn{1}{|c|}{Group} &
 \multicolumn{1}{|c}{[350]} &
 \multicolumn{1}{c}{[450]} &
 \multicolumn{1}{c}{[750]} &
 \multicolumn{1}{c}{[800]} &
 \multicolumn{1}{c}{[850]} &
 \multicolumn{1}{c}{[1100]} &
 \multicolumn{1}{c}{[1300]} &
 \multicolumn{1}{c|}{[2700]} &
 \multicolumn{1}{|c|}{M$_{disk} [M_\odot]$} &
 \multicolumn{1}{|c|}{$n$} & 
 \multicolumn{1}{|c|}{$\sigma_{n}$} & 
 \multicolumn{1}{|c|}{Ref.} \\ 
 (1)                      &(2)&  (3) & (4)   & (5)  & (6)  & (7)   & (8)  & (9) & (10)                 &  (11)               &  (12)  &  (13)  & (14)   \\
 \hline 
V376 Cas                  & I & 5.8$\pm$0.5 & 1.3$\pm$0.2  & 0.28$\pm$0.04& 0.23$\pm$0.02&       &0.08$\pm$0.02 & 0.038$\pm$0.006& $<$0.0048           &8.8 $\times 10^{-2}$ &  4.59  &  0.21  & 3, 7, 13 \\
Elias 3-1$^\flat$         & I & 4.1$\pm$0.6 & 2.6$\pm$0.4  & 0.8$\pm$0.1& 0.78$\pm$0.05& 0.64$\pm$0.05 &0.35$\pm$0.05 & 0.23$\pm$0.02& 0.05$\pm$0.01  	       &3.5 $\times 10^{-2}$ &  3.22  &  0.03  & 1, 3, 7  \\
AB Aur                    & I & 9$\pm$1 & 3.8$\pm$0.6  & 0.53$\pm$0.08& 0.50$\pm$0.05& 0.36$\pm$0.07 &0.15$\pm$0.02 & 0.10$\pm$0.02& 0.011$\pm$0.005 	       &1.2 $\times 10^{-2}$ &  4.32  &  0.09  & 3, 7, 8  \\
HD 34282$^\bigstar$       & I &    &1.6$\pm$0.2$^\sharp$&  & 0.41$\pm$0.03& 0.38$\pm$0.02 &0.18$\pm$0.02 & 0.11$\pm$0.01& 0.024$\pm$0.003$^b$ 	       &1.0 $\times 10^{-1}$ &  3.36  &  0.04  & 9, 14, 15, 17  \\
HD 34700                  & I &      & 0.22$\pm$0.04  &      &      & 0.041$\pm$0.002 &0.04$\pm$0.01 & 0.018$\pm$0.002&        	       &1.2 $\times 10^{-2}$ &  3.64  &  0.34  & 15, 17, 19  \\
HD 97048                  & I &      &       &      &      &       &      & 0.45$\pm$0.03&                     &9.2 $\times 10^{-2}$ &$>$2.94 &        & 4 \\
HD 100453                 & I &      &       &      &      &       &      & 0.27$\pm$0.02&                     &2.1 $\times 10^{-2}$ &$>$2.82 &        & 10 \\ 
HD 100546                 & I &      &       &      &      &       &      & 0.47$\pm$0.02& 0.036$\pm$0.004$^c$           &2.9 $\times 10^{-2}$ &  4.20  &  0.50  & 5, 18  \\
HD 135344                 & I & 6$\pm$2  & 3.2$\pm$0.2  &      & 0.57$\pm$0.02& 0.49$\pm$0.01 &0.21$\pm$0.01 & 0.14$\pm$0.02& $<$0.076$^a$        &1.6 $\times 10^{-2}$ &  3.89  &  0.05  & 2, 15, 16  \\
HD 139614                 & I &      &       &      & 0.61$\pm$0.03&       &0.27$\pm$0.01& 0.24$\pm$0.01& 0.08$\pm$0.01$^a$           &2.7 $\times 10^{-2}$ &  3.17  &  0.19  & 15   \\
HD 142527                 & I &      &       &      & 4.2$\pm$0.5&       &      & 1.19$\pm$0.03& 0.047$\pm$0.006$^c$           &1.5 $\times 10^{-1}$ &  3.58  &  0.41  & 18, 19  \\
HD 169142                 & I &      & 3.3$\pm$0.1  &      & 0.55$\pm$0.03& 0.57$\pm$0.01 &0.29$\pm$0.01 & 0.20$\pm$0.02& 0.07$\pm$0.02$^a$           &2.4 $\times 10^{-2}$ &  3.56  &  0.11  & 15, 16  \\
HD 179218                 & I &      &       &      &      &       &      & 0.071$\pm$0.007& 0.008$\pm$0.001  	       &2.4 $\times 10^{-2}$ &  4.06  &  0.26  & 9  \\
 \hline		    	       			           
\bf{Mean}                 & \bf{I} & &       &      &      &       &      &      &                     &                    &\bf{3.60}&\bf{0.09}&   \\
 \hline		    			       	   
HD 31648                  & II&      & 3.3$\pm$0.3  &      &      & 0.78$\pm$0.02 &      & 0.36$\pm$0.02& 0.04$\pm$0.01 	       &3.6 $\times 10^{-2}$ &  3.44  &   0.18 & 8, 16  \\
UX Ori$^\bigstar$         & II&      &       &      &      &       &      & 0.020$\pm$0.002& 0.0038$\pm$0.0004$^b$          &1.6 $\times 10^{-2}$ &  3.20  &   0.29 & 12  \\
HD 35187                  & II&      & 0.27$\pm$0.05  &      & 0.12$\pm$0.02& 0.06$\pm$0.005 &0.08$\pm$0.01 & 0.035$\pm$0.002&        	       &5.0 $\times 10^{-3}$ &  2.77  &   0.38 & 15, 17, 19 \\
BF Ori$^\bigstar$         & II&      &       &      &      &       &      & 0.006$\pm$0.002&                     &1.0 $\times 10^{-2}$ &$>$2.82 &        & 11     \\
HD 104237                 & II&      &       &      &      &       &      & 0.09$\pm$0.02&                     &7.8 $\times 10^{-3}$ &$>$2.79 &        & 5   \\
HD 141569                 & II&      & 0.07$\pm$0.02 &      &      & 0.014$\pm$0.002 &$<$0.036& 0.009$\pm$0.004&                   &5.6 $\times 10^{-4}$ &  2.92  &   0.38 & 15, 17, 19  \\
HD 142666$^\bigstar$      & II&      & 1.14$\pm$0.04  &      & 0.35$\pm$0.02& 0.313$\pm$0.005 &0.18$\pm$0.01 & 0.127$\pm$0.09& $<$0.063$^a$        &1.6 $\times 10^{-2}$ &  3.07  &   0.01 & 15, 16  \\
HD 144432                 & II&      &       &      & 0.10$\pm$0.03&       &0.07$\pm$0.01 & 0.037$\pm$0.003&             	       &4.9 $\times 10^{-3}$ &  3.00  &   0.62 & 15, 19  \\
HD 150193                 & II&      & 0.5$\pm$0.2 &      &      & 0.12$\pm$0.02 &      & 0.05$\pm$0.01& 0.014$\pm$0.003  	       &6.3 $\times 10^{-3}$ &  2.97  &   0.11 & 6, 8 \\
AK Sco$^\bigstar$         & II&      & 0.24$\pm$0.03  &      &      & 0.083$\pm$0.009 &0.036$\pm$0.004 &      &            &3.1 $\times 10^{-3}$$^\dagger$&  3.05  &   0.35 & 6  \\
HD 163296$^\flat$         & II&10.7$\pm$0.7 & 5.6$\pm$0.4  & 2.65$\pm$0.08& 2.32$\pm$0.04& 1.92$\pm$0.03 &1.02$\pm$0.05 & 0.78$\pm$0.03&         	       &6.8 $\times 10^{-2}$ &  2.93  &   0.09 & 7  \\
WW Vul$^\bigstar$         & II&      &       &      &      &       &      & 0.011$\pm$0.001&                     &1.3 $\times 10^{-2}$ &$>$2.61 &        & 11  \\
SV Cep$^\bigstar$         & II&      &       &      &      &       &      & 0.007$\pm$0.002&                     &8.6 $\times 10^{-3}$ &$>$3.18 &        & 11  \\
 \hline		      	       			           
\bf{Mean}                 &\bf{II}&  &       &      &      &       &      &      &                     &                    &\bf{3.06}&\bf{0.06}&     \\
 \hline	
\end{tabular}		

\begin{tabular}{|cl|cl|cl|}
 \hline
 \multicolumn{6}{|c|}{References}\\ \hline \hline
1 & Beckwith \& Sargent (1991)  &8 & Mannings \& Sargent (1997)  &15& Sylvester et al. (1996)                 \\		             
2 & Coulson \& Walther (1995)   &9 & Mannings \& Sargent (2000)  &16& Sandell \& Weintraub, AAS Poster (2003) \\      
3 & di Francesco et al. (1997)  &10& Meeus et al. (2003)         &17& Sheret et al. (2004)                    \\         
4 & Henning et al. (1993)       &11& Natta et al. (1997)         &18& van den Ancker (unpublished ATCA data)  \\      
5 & Henning et al. (1994)       &12& Natta et al. (1999)         &19& Walker \& Butner (1995)                 \\                   
6 & Jensen et al. (1996)        &13& Osterloh \& Beckwith (1995) &  &        \\                                        
7 & Mannings (1994)             &14& Pi\'etu et al. (2003)       &  &        \\                                       
 \hline
\end{tabular} 
\end{center}	
\normalfont  
\end{sidewaystable*}		  

\section{Discussion and Interpretations \label{interp}}

\subsection{Comparison with other articles}

Hillenbrand et al. 1992 (henceforth H92) also classified HAEBE stars
in groups. We stress that this was performed based on the
\textit{near-IR} ($\sim$3.5--10.6~$\mu$m) spectral
index. Their classification is not linked to the M01 classification
used in this paper; both H92 group I and II contain M01 group I as well as
group II sources. H92 group III is not relevant for sources with
significant dust disks. The classification of M01 focusses on the 
$\sim$20--100~$\mu$m SED, which makes it well-suited for
this observational analysis.

There exists a large variety of models for the SEDs of HAEBEs in the
literature (e.g. 
H92; Natta et al. 1992, 1993; Mannings 1994). While most of these
(disk and/or envelope)
models give satisfactory fits to the mid-IR SED, they have difficulty
to accurately 
describe the near-IR bump that is typical for HAEBEs. One of the main
improvements in the model of Dullemond et al. (2001) and DD04 is the
inclusion of the puffed-up inner rim, which generates this near-IR
bump, as was first suggested by Natta et al. (2001) and Tuthill et
al. (2001). Furthermore,
the Dullemond et al. model explains in a natural way the roughly two types of IR
SEDs (M01 group I and II) observed in HAEBE stars. 
The quantitative fine-tuned modelling of the SEDs of our sample
stars, however, is beyond the scope of the present observational
article.

\subsection{Grain growth in circumstellar disks around HAEBEs \label{gg}}

If the disks surrounding HAEBE stars are optically thin at \submm
wavelengths, the spectral index of the SED in this wavelength range
will be independent of disk geometry. In that case, intrinsic properties
of the cold grains in the outer parts of the disk must be responsible
for the observed variations in the \submm spectral index.

In order to test the hypothesis that the circumstellar disks in our
sample are indeed optically thin at \submm wavelengths,
we have computed the radii of the circumstellar disks of our sample
sources based on the measured 1.3 mm flux, assuming the simple model
of an optically \textit{thick} disk with a black-body
temperature of 40K. The resulting radii are of the order of $\sim$10--20
AU. Comparing this to the observed radii of Herbig Ae stars (e.g. Mannings \& Sargent
1997, Natta et al. 2003, Testi et al. 2003, Pi\'etu et al. 2003) or
the typical disk sizes of T Tauri stars 
($\sim$100 AU), the estimated radii of our sample-star disks seem too small. It
is therefore unlikely that the disks of our sample stars are optically
thick at \submm wavelengths.

Supposing an optically \textit{thin} disk at \submm wavelengths and a
typical grain temperature of 40K, one can estimate the disk mass from
the 1.3 mm flux. Using formula (14) in Hillenbrand et al. (1992), we
derived disk masses for the sample sources
between $5.6 \times 10^{-4}$ and $1.5 \times 10^{-1}$ M$_\odot$. This
compares well to the maximum 
disk mass of 0.3~M$_\star$, set by gravitational instabilities in the
disk (e.g. Gammie 2001). When the disk is optically thick, its mass
would be several factors higher, which in some cases (e.g. HD 34282)
would be close to this theoretical limit.

We have determined the value of the slope of the \submm SED by means
of a power law fitted to the data between 350 and 2700~$\mu$m. It has
not been proven that Herbig Ae/Be disks are still optically thin at
the shortest wavelengths of this interval. 
However, for the sources for which we have measurements at the shorter
wavelengths, we can also compute the \submm slope based on the
longer-wavelength data only. There is no significant difference
between the two. Hence, the slope of the energy distribution down to
350 and 450 micron does not deviate significantly from a power law. 
This shows that the optical depth at these wavelengths is not very
different from that at 1.3 mm. Another way of saying this is that our
results would not change when omitting the 350 and 450 micron data.

Laboratory studies of the emission properties of dust
species with astronomical relevance (e.g. Mennella et al. 1998;
Colangeli et al. 2001) show that variations in chemical composition
only exert a very 
mild influence on the \submm emission characteristics of dust
particles. Particle size, on the other hand, has quite a large
influence on the \submm emission. Therefore, and in accordance with
previous authors (e.g. Mannings \& Emerson 1994; van den Ancker et al. 2000;
Bouwman et al. 2000; Testi et al., 2003),
we attribute the differences in \submm-spectral-energy-distribution
slope to variations in dust particle size.

\begin{figure}
\rotatebox{0}{\resizebox{3.5in}{!}{\includegraphics{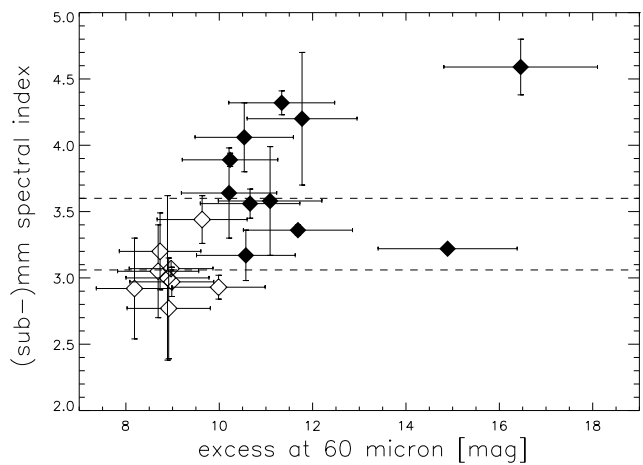}}}
\caption{ The \submm spectral index versus the flux excess at 60
  micron in magnitudes. Group I sources are represented by filled, 
  group II by open symbols. The upper dashed line represents the
  average value of $n$ in group I, the lower line the average in group
  II.}
\label{excvsn.ps}
\end{figure}

It is clear from Fig.~\ref{plotroy.ps} and Table~\ref{grIorIIvsn} that
the \submm spectral indices and the classification of the sample stars
are correlated. Group I sources have an average spectral index $n$ equal to
$3.60\pm0.09$. In group II, $n$ is $3.06\pm0.06$. 
Fig.~\ref{excvsn.ps} describes this correlation alternatively.
The spectral index is plotted versus the flux excess at 60 micron.
The latter quantity is the difference between the
measured flux and the flux of the reddened Kurucz model (see
Fig.~\ref{seds.ps}) at this wavelength.
Group I objects, which are by definition brighter
mid-IR sources, indeed display higher values for $n$. 
We interpret this
correlation in terms of a correlation between the grain size
distribution of the cold grains and the disk geometry. Group I sources
(flared-disk systems) appear to have smaller grains than
group II sources (self-shadowed-disk objects).

Two explanations are given here.
Due to coagulation, dust particles grow. It is therefore possible that
different grain sizes indicate different evolutionary stages.
Group I sources have dust that is only marginally evolved compared to
ISM dust. Group II sources on the other hand have a significant
population of mm-sized grains. We suggest
that flared disks, in which coagulation takes place, evolve into
geometrically somewhat flatter
self-shadowed disks (by dust settling to the midplane) and become group
II members. 

Alternatively, it is possible that, in contrast to coagulation, there are
mechanisms that \textit{replenish} the small-sized-grain population in
circumstellar disks. The presence of the larger particles in self-shadowed-disk
systems would then imply that this replenishment is not active/efficient
(anymore) in these sources. In this hypothesis, flared disks are not necessarily
evolutionally connected to self-shadowed disks, and the geometry of
the disk is kept flared by the replenishment of the small grains.

We note that the shape of the amorphous 10 micron silicate feature,
which is linked to the particle sizes of the warm silicates (Bouwman
et al. 2001, van Boekel et al. 2003), does not give any direct
information 
about the grain size distribution of the silicates in the
mid-plane of the circumstellar disk. This feature is believed to
originate from the surface layers of the
disk. Grain growth of these warm grains has been observed in this
sample of HAEBEs, but is uncorrelated with the \submm spectral index
(see AV04). Hence these small ($\sim$ 0.1--2 $\mu$m) warm grains seem to
have an evolution, independent of the grain growth in the interior of
the disk. On the other hand, the 10 micron feature can be an
indication for the sizes of the smallest grains in the disk, assuming
that efficient vertical mixing in the disk occurs.

\section{Conclusions \label{concl}}

We showed in this paper that there is a significant
difference between the \submm spectral indices of HAEBE stars with
different IR SEDs. This could very well mean that the grain
size distribution of the cold grains in a HAEBE star depends
on the disk geometry.

The HAEBE systems with large IR excesses (group I, i.e. flaring-disk
objects) have smaller cold grains than systems with more 
moderate IR excesses (group II, i.e. self-shadowed-disk sources). 
In other words, it appears that \textit{the flaring geometry of a
circumstellar disk can only persist when a sufficiently large
population of small grains exists in the disk}.

The circumstellar-disk models of DD04 agree qualitatively with the
results presented here. Disks in which the vast majority of the grains
have grown to larger sizes, lead automatically to self-shadowed
geometries. Models with small grains result in a mid-infrared SED
akin to group I sources.
Models with grain sizes larger than $\sim$1 mm not only display 
mid-infrared SEDs more resembling those of group II sources, but also
change their \submm slope from larger values of 
$n$ to a black-body character ($n=3$). DD04 note that the \submm
slope changes with somewhat smaller grain sizes than are required to
alter the mid-IR SED from group I to group II. 
Therefore, according to the models, 
the spectral index can have all values larger than 3 in group I
sources, and all group II sources should have a black-body \submm slope.
This is consistent with the observed trend in which group I sources
are observed to have $3.17<n<4.59$ and group II sources have $n$ very
close to 3. The DD04 models predict that grain growth resulting in a 
change of $n$ from values similar to those found in the ISM ($> 3$) to
the optically thin black-body value ($n=3$) occurs before the disk
becomes self-shadowed. However the DD04 models do not model the
dynamical processes leading to grain growth. At present the time scale
of this transition from larger $n$ to $n=3$ is unknown, 
so we are unable to estimate the fraction of group I sources that
should have $n=3$. Therefore we have no reason to assume a priori
that our distribution of (sub-)mm-spectral-index values is
incompatible with the interpretation outlined in Section~\ref{gg}.

In the present paper, we report a clear correlation between the shape
of the mid-IR SED and 
the \submm slope in our sample. Previous authors have shown that
variations in \submm slope within systems with proto-planetary disks
are caused by variations in dust particle size. D02 have shown that
the shape of the mid-IR SED is a proxy for disk geometry (flared
versus self-shadowed). In accordance with these results, we interpret our find
as a correlation between disk geometry and grain size. 
If one assumes that in the course of disk evolution grains will grow,
this offers the first observational indications to the idea that
flaring disks evolve into self-shadowed disks.


\begin{thebibliography}{}

\bibitem[]{} Acke, B., van den Ancker, M.E., 2004, A\&A, submitted
\bibitem[]{} Beckwith, S.V.W., Sargent, A.I., 1991, ApJ 381, 250
\bibitem[]{} Bouwman, J., de Koter, A., van den Ancker, M.E., Waters,
  L.B.F.M., 2000, A\&A 360, 213 
\bibitem[]{} Bouwman, J., Meeus, G., de Koter, A., et al. 2001, A\&A 375, 950 
\bibitem[]{} Chiang, E.I., Goldreich, P., 1997, ApJ 490, 368
\bibitem[]{} Colangeli, L., Battaglia, R., Brucato, J. R., Bussoletti,
  E., della Corte, V., Esposito, F., Ferrini, G., Mazzotta Epifani,
  E., Menella, V., Palomba, E., Palumbo, P., Rotundi, A., 2001,
  Meteoritics \& Planetary Science 36, 44
\bibitem[]{} Coulson, I.M., Walther, D.M., 1995, MNRAS 274, 977
\bibitem[]{} di Francesco, J., Evans, N.J., II, Harvey, P.M.; Mundy, L.G., 
Guilloteau, S., Chandler, C., 1997, ApJ 482, 433
\bibitem[]{} Dullemond, C.P., Dominik, C., Natta, A., 2001, ApJ 560, 957
\bibitem[]{} Dullemond, C.P., 2002, A\&A 395, 853
\bibitem[]{} Dullemond, C.P., van den Ancker, M.E., Acke, B., van Boekel, R.,
2003, ApJ 594, 47
\bibitem[]{} Dullemond, C.P., Dominik, C., 2004, astro-ph/0401495
  (A\&A, in press)
\bibitem[]{} Egan, M.P., Price, S.D., Shipman, R.F., Tedesco, E., 1997, AAS 191,
5008
\bibitem[]{} Gammie, C.F., 2001, ApJ 553, 174
\bibitem[]{} Henning, T., Pfau, W., Zinnecker, H., Prusti, T., 1993, A\&A 276, 129
\bibitem[]{} Henning, T., Launhardt, R., Steinacker, J., Thamm, E., 1994, A\&A 291, 546
\bibitem[]{} Henning, T., Burkert, A., Launhardt, R., Leinert, C.,
  Stecklum, B., 1998, A\&A 336, 565
\bibitem[]{} Hillenbrand, L.A., Strom, S.E., Vrba, F.J., Keene, J., 1992, ApJ 397, 613
\bibitem[]{} Jensen, E.L.N., Mathieu, R.D., Fuller, G.A., 1996, ApJ 458, 312
\bibitem[]{} Malfait, K., Bogaert, E., Waelkens, C., 1998, A\&A 331, 211
\bibitem[]{} Mannings, V., 1994, MNRAS 271, 587
\bibitem[]{} Mannings, V., Emerson, J.P., 1994, MNRAS 267, 361
\bibitem[]{} Mannings, V., Sargent, A.I., 1997, ApJ 490, 792
\bibitem[]{} Mannings, V., Sargent, A.I., 2000, ApJ 529, 391
\bibitem[]{} Meeus, G., Waters, L.B.F.M., Bouwman, J., van den Ancker, M.E.,
Waelkens, C., Malfait, K., 2001, A\&A 365, 476
\bibitem[]{} Meeus G., Bouwman J., Dominik C., Waters L.B.F.M., de Koter A., 2003, 
 A\&A 402, 767
\bibitem[]{} Mennella, V., Brucato, J.R., Colangeli, L., Palumbo, P.,
  Rotundi, A., Bussoletti, E., 1998, ApJ 496, 1058 
\bibitem[]{} Natta, A., Palla, F., Butner, H.M., Evans, N.J., Harvey, P.M., 1992,
ApJ 391, 805
\bibitem[]{} Natta, A., Prusti, T., Kr\"ugel, E., 1993, A\&A 275, 527
\bibitem[]{} Natta A., Grinin V.P., Mannings V., Ungerechts H., 1997, ApJ 491, 885
\bibitem[]{} Natta, A., Prusti, T., Neri, R., Thi, W.F., Grinin, V.P., Mannings, V.,
1999, A\&A 350, 541
\bibitem[]{} Natta A., Grinin, V.P., Mannings, V., 2000, Protostars and Planets IV, 559
\bibitem[]{} Natta, A., Prusti, T., Neri, R., et al. 2001, A\&A 371, 186
\bibitem[]{} Natta, A., Testi, L., Neri, R., Shepherd, D.S., Wilner, D.J., 2003,
A\&A, in press
\bibitem[]{} Osterloh, M., Beckwith, S.V.W., 1995, ApJ 439, 288
\bibitem[]{} Pi\'etu, V., Dutrey, A., Kahane, C., 2003, A\&A 398, 565
\bibitem[]{} Sheret, I., Dent, W.R.F., Wyatt, M.C., 2004, MNRAS 348, 1282
\bibitem[]{} Sylvester, R.J., Skinner, C.J., Barlow, M.J., Mannings, V., 1996, MNRAS 279, 915
\bibitem[]{} Testi, L., Natta, A., Shepherd, D.S., Wilner, D.J., 2003, A\&A 403, 323
\bibitem[]{} Th\'e, P.S., de Winter, D., P\'erez, M. R., 1994, A\&AS 104, 315
\bibitem[]{} Tuthill, P.G., Monnier, J.C., Danchi, W.C., 2001, Nature 409, 1012
\bibitem[]{} van Boekel, R., Waters, L.B.F.M., Dominik, C., Bouwman, J., de
Koter, A., Dullemond, C.P., Paresce, F., 2003, A\&A 400, 21
\bibitem[]{} van den Ancker, M.E., Bouwman, J., Wesselius, P.R., Waters,
L.B.F.M., Dougherty, S.M., van Dishoeck, E.F., 2000, A\&A 357, 325
\bibitem[]{} Walker H., Butner H., 1995, Ap\&SS 224, 389

\end{thebibliography}
\end{document}